Invited review article

**Title:** Bioinspired approaches to toughen calcium phosphate-based ceramics for bone repair.

**Authors:** Peifang Dee[1,2], Ha Young You[3], Swee-Hin Teoh[4,5], Hortense Le Ferrand[1,2,3*]

**Addresses:**
[1]School of Mechanical and Aerospace Engineering, Nanyang Technological University, 50 Nanyang Avenue, Singapore 639798
[2]Singapore Centre for 3D Printing, 50 Nanyang Avenue, Singapore 639798
[3]School of Materials Science and Engineering, Nanyang Technological University, 50 Nanyang Avenue, Singapore 639798
[4]School of Chemical and Biomedical Engineering, Nanyang Technological University, 62 Nanyang Drive, Singapore 637459
[5]Lee Kong Chian School of Medicine, Nanyang Technological University, 11 Mandalay Road, Singapore 308232

**Corresponding author:**
*hortense@ntu.edu.sg

**Highlights:**
- Bone has a complex multi-level hierarchical organisation. This structural organisation is key for its outstanding properties.
- Artificial calcium phosphate ceramics (CPCs) are promising for achieving the set of mechanical and biological requirements for use as bone implants.
- At individual lengthscales, methods have been developed to control crystal phases, microscale porosity and macroscopic geometry but the CPCs produced remain intrinsically brittle.
- Inspired by the organisation of bones and other biomaterials, CPCs and their composites could have their toughness increased through multi-level hierarchical microstructures.
- Challenges remain but recent advances in ceramic processing and toughening may pave the way for a new generation of medical implants.

*Abstract*

*To respond to the increasing need for bone repair strategies, various types of biomaterials have been developed. Among those, calcium phosphate ceramics (CPCs) are promising since they possess a chemical composition similar to that of bones. To be suitable for implants, CPCs need to fulfill a number of biological and mechanical requirements. Fatigue resistance and toughness are two key mechanical properties that are still challenging to obtain in CPCs. This paper thus reviews and discusses current progress in the processing of CPCs with bioinspired microstructures for load-bearing applications. First, methods to obtain CPCs with bioinspired structure at individual lengthscales, namely nano-, micro-, and macroscale are discussed. Then, approaches to attain synergetic contribution of all lengthscales through a complex and biomimetic hierarchical structure are reviewed. The*

*processing methods and their design capabilities are presented and the mechanical properties of the materials they can produce are analysed. Their limitations and challenges are finally discussed to suggest new directions for the fabrication of biomimetic bone implants with satisfactory properties. The paper could help biomedical researchers, materials scientists and engineers to join forces to create the next generation of bone implants.*

# 1| Introduction

Bone autografts are the gold standard for bone repair. With the ageing population, changes in diet, the global rise in diabetes and other health problems have been weakening our skeleton, leading to an increasing prevalence of dramatic bone injuries or amputations. Many strategies have thus been developed to help patients recover their limbs. These strategies include allogenic grafts, tissue engineering approaches, and implantation of bioinert and bioactive materials. Allogenic grafts have limited supply and have a potential risk of inflammation and disease transmission. Tissue engineering (TE) approaches are time-consuming since cells are directed to re-mineralize bone in response to the signals sent by an implanted porous materials. Allographs and TE approaches are thus non-ideal solutions for the repair of load-bearing bones whereby a quick recovery of the structural functions is required. As an alternative, bioinert implants made from cement, ceramics or metals are currently clinically used. However, in practice, those implants tend to fail after 10-15 years. Indeed, insufficient bioadhesion to the surrounding tissues and unmatching properties with the native bone often lead to micromotion, wear, inflammation, as well as stress shielding and bone decay around the implant [1]. Bioactive implants, in turn, are materials with a chemical composition close to that of bones. They provide good osseointegration but are often limited in their mechanical performance. This paper will be discussing this last type of artificial implants.

The main mineral component of bone is calcium phosphate (CaP). Bioactive implants thus contain calcium under the form of glass [2], calcium-based cements [3] and calcium phosphate ceramics (CPCs), composites and coatings. Among the various CPCs, hydroxyapatite (HA), tricalcium phosphate (TCP), and a mixture of both, biphasic calcium phosphate (BCP), have been the most studied and the most promising candidates for bone repair [4]. Indeed, these CPCs fulfill most of the most important criteria required for their use as load-bearing clinical implants, which are [5]:

   (i)     To present bone-matching "static" mechanical properties, such as stiffness, hardness and strength.
   (ii)    To promote osseointegration and remodelling processes.
   (iii)   To have a similar weight as the original bone defect.
   (iv)    To be shaped according to the defect.
   (v)     To be available quickly and without limitation of material.
   (vi)    To present "dynamic" bone-like mechanical properties, such as fatigue cycle resistance, stable crack propagation, and toughness.

However, this last criterion (vi), namely toughness and fatigue resistance, is still particularly challenging to obtain in bioactive ceramics **(Figure 1)**. The comparison of the mechanical properties of most common bioceramics with those of cancellous (porous or trabecular) and cortical (dense or compact) bones, indicates that although the Young's modulus and flexural strength of CPCs are generally satisfactory (**Figure 1A,B**), their toughness and resistance to fatigue are still mediocre **(Figure 1C)**. Since synthetic materials do ont self-repair like biomaterials do, damage tolerant properties with flexural fracture toughness are very important properties to pursue. It is therefore critical to pursue research to engineer biomaterials with such mechanical properties.

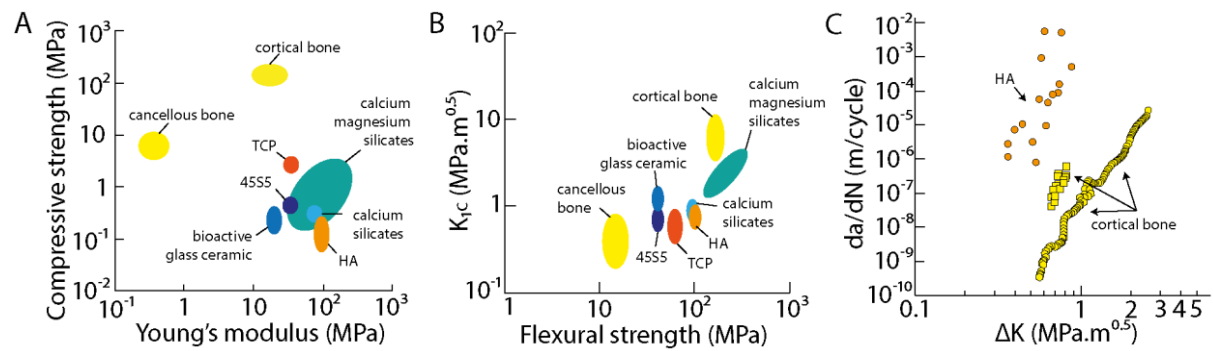

**Figure 1: Mechanical properties of human bones and most common bioceramics: (A)** Compressive strength as a function of the Young's modulus, **(B)** toughness $K_{1c}$ as a function of the flexural strength and **(C)** fatigue-crack growth rate as a function of the stress intensity range $\Delta K$ for hydrated human bones and a representative bioceramics. Data are extracted from refs [6–8]. $K_{1c}$ indeed indicates the energy required to initiate a crack, whereas $\Delta K$ is the increase in energy required to propagate this crack. 45S5 is a commercially available bioactive glass.

Toughening mechanisms in natural load-bearing bones have been studied extensively [9–12]. Healthy long bones display high toughness in their transverse direction. This is illustrated by a rising resistance curve, or R-curve, where the stress intensity factor $K_{jc}$ increases with the crack length (**Figure 2A**) [10]. Indeed, hydrated bones loaded transversally do not break in a catastrophic brittle manner. Instead, extrinsic toughening mechanisms such as microcracking, fiber pull-out, crack deflections and crack twisting are observed in the cortical part (**Figure 2B-D**) [10,13]. These mechanisms are intimately related to the intricate hierarchical organisation of bones, spanning across 12 lengthscales (**Figure 2E**) [14].

Developing strategies to toughen structural materials has been the focus of many excellent reviews. In the context of CaP materials, there has been many excellent reviews tackling the mechanics of CaP composites [15], their bioactivity [16] and the additive manufacturing of ceramic scaffolds [17]. This paper aims at complementing the existing literature by providing an overview of CPCs manufacturing processes that allow multiscale structural control to better reproduce bone architecture and properties. Indeed, bioinspired approaches have been studied for many years and have led to outstanding results in other materials [18]. First, we present processing methods of CPCs with structural control at individual lengthscales, namely at the nanoscale with the control of crystalline phases and their transformation, at the microscale with the formation of intended micropores, and at the macroscale with the design of complex shapes. For each lengthscale, the methods are described along with their structural design capabilities and their impacts on the mechanical and biological properties. Then, multiscale bioinspired approaches explored to increase the mechanical performance of those materials are reviewed. Those approaches comprise the use of reinforcing particles, and the construction of hierarchical features inspired by bones and other biomaterials. Finally, the challenges and limitations in the fabrication of CPCs for bone repair are discussed. Future research directions are suggested based on the recent advances in processing of toughened ceramics. The fabrication techniques and resulting properties presented in this review highlight the need for collaborations between material scientists, mechanical engineers, chemists, biologists and clinicians. The results reviewed here could constitute a common basic knowledge and understanding across the disciplines and suggest ideas for future research towards improved bone implants.

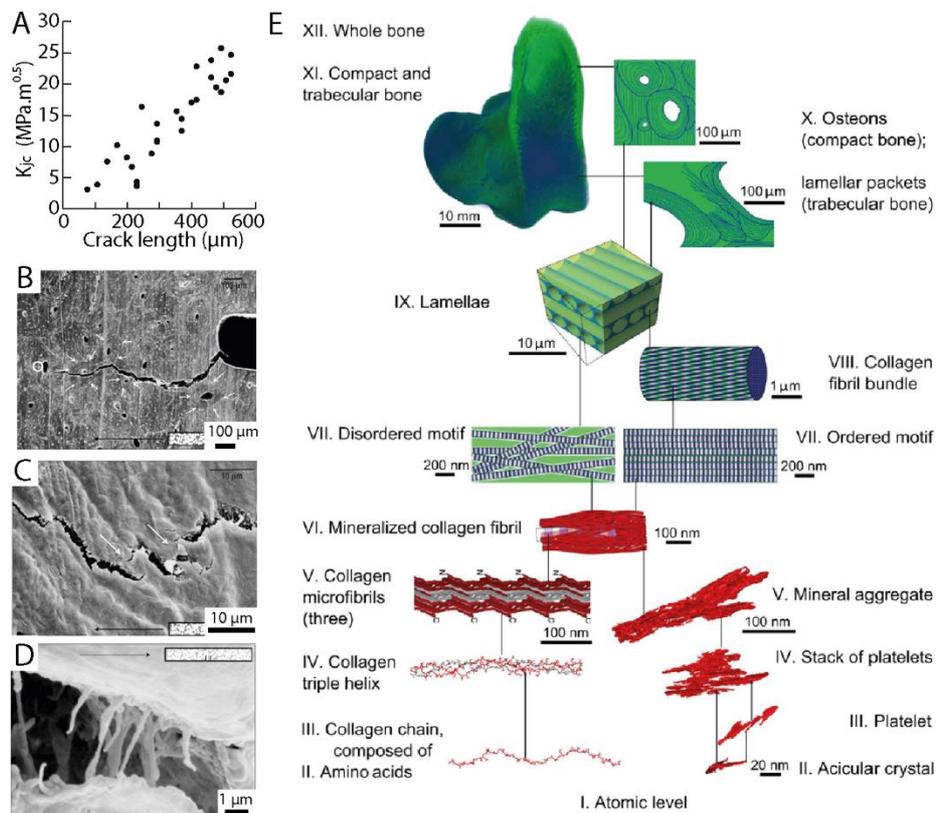

**Figure 2: Multi-level hierarchical structure of bones: (A)** Stress intensity factor $K_{jc}$ as a function of the crack length for a hydrated human long bone loaded transversally. Data extracted from ref [10]. **(B-D)** Scanning electron micrographs showing crack paths in human cortical bones loaded transversally. Reprinted from ref [13] with permission from Springer Nature. **(E)** Schematics showing the 12 hierarchical levels in bone structure. From ref [14]. Reprinted with permission from AAAS.

## 2| Bioinspired approaches at individual lengthscales

In the following we review the bioinspired approaches and resulting mechanical properties used to engineer CPCs with structural control at the nano-, micro-, and macroscale. At the nanoscale, crystalline phases and grain sizes are key for satisfying static mechanical (requirement (i)) and biological properties (requirement (ii)). At the microscale, the porosity plays an important role for the osseointegration (requirement (ii)) and lightweightness (requirement (iii)). At the macroscale, the scalability, speed, and complexity of the parts can be controlled (requirements (iv) and (v)). Since the primary focus of the paper is on long load-bearing bones, the properties of the fabricated materials are compared with both cortical and cancellous bone.

### 2.1| Nanoscale: control of the CaP crystalline phase

The main CaP phase present in bones consists in nanoscopic hydroxyapatite fibers and platelets. During the fabrication of CPCs, a calcium-rich powder is prepared, packed or assembled, and sintered at high temperature for consolidation. During the heat treatment, CPCs are subject to phase transformations. In this section, we introduce those phase transformations, the effects of using biomimetic and waste CaP nanopowders as raw materials, and discuss the properties of the sintered CPCs (**Figure 3**).

Understanding the phase transformations of CPCs is important to prevent the development of internal stresses and defects, and to obtain the expected crystalline phase after sintering that may impact the biological response. However, the phase diagram of CPCs is complex and depends on a large number of parameters, such as temperature, pressure, composition, atmosphere **(Figure 3A)**. For example, controlling the water vapor during sintering plays can stabilize certain calcium phosphate phases [19]. In practice, HA, TCP or mixtures of powders are densified by sintering in air at temperatures ranging from 1000 to 1250 °C, with or without pressure [20–22]. HA can remain stable up to 1350-1450 °C in air but partial dehydration may occur, ultimately leading to residual pores and cracks [23]. Calcium-deficient HA (CDHA) has been used instead of HA for its high sintering rate, but CDHA decomposes at 700 °C into BCP [24]. TCP also undergoes phase transformations. One of them has a dramatic effect: the transformation of $\alpha$-TCP into β-TCP at ~1120 °C that is accompanied by a crystal size contraction that generates cracks [25]. Pure HA ceramics sintered at 1100-1250 °C are ~98% dense and feature submicrometric grains of 1 to 10 μm, leading to a compressive strength of 40 MPa and a toughness of 1-1.4 MPa.m$^{0.5}$. Substituting HA with ions such as Zn, Ni, Sr and Mn was found to aid the sintering but leads to thermal decomposition and BCP phases without improvement in toughness [26–28]. An increase in toughness to 2.7 MPa.m$^{0.5}$ could be nevertheless achieved by sintering HA powders in inert atmosphere [29] or by adding fluoroapatite to retain small nanometric grains after sintering [30]. Among the many studies reporting the sintering of CPCs, there is therefore little success in obtaining CPCs with a toughness and crack propagation resistance required for bone implants.

One proposition to improve CPCs' mechanical and biological properties has been to use powders with crystal phases and grain sizes identical to that of bone [31]. Indeed, synthetic HA crystals, despite showing osseointegration, are still considered as foreign material by the body [32]. Although the crystalline phases, orientations and sizes of bone crystals remain unclear, with a mixture of carbonate-substituted HA nanocrystals, polycrystals and other disordered phases [14], several strategies have been developed to synthesize biomimetic CaP nanoparticles. Syntheses carried out in presence of biomolecules such as collagen, bovine serum albumin or urea have been the most promising to create biomimetic nanocrystals [33–37]. Those nanocrystals are carbonated HA with needle-like shape and lengths varying from 27 nm to 1.5 μm. Sintering compacts made of these nanoparticles at 1200-1300 °C in air showed densification up to 97% and grain sizes of 1-3 μm, resulting in bending strengths of ~88 MPa, close to that of bones [36,37]. However, toughness has not been reported.

Another approach to yield biomimetic crystalline phases in scalable amounts is to use waste from bones [38] or from calcined eggshells. Waste bone powders are first reacted with a phosphate source such as orthophosphoric acid or dicalcium hydrogen phosphate before being compacted and sintered at 1100 °C [39,40]. Ceramics with ~47% porosity were obtained, but they featured pure HA crystals and nanosized grains of ~111 nm [38]. Despite the porosity that lead to a low hardness, compression strength and modulus, of 1 GPa, 0.9

MPa and 6 GPa, respectively, the toughness was higher than bulk synthetic hydroxyapatite with a $K_{1c}$ of 2.21 MPa.m$^{0.5}$ [38]. This toughness has been attributed to the presence of uniformly distributed round nanopores that could act as crack deflectors (**Figure 3B**). As a comparison, the sintering of the powders synthesized from eggshells could be densified to 98% but the grain obtained had sizes above 1 µm, thereby resulting in a high hardness of 5.5-6 GPa but a toughness of 1-1.5 MPa.m$^{0.5}$ only.

The properties of CPCs made from synthetic, biomimetic and biowaste powders are related to the initial powder sizes and phases, and to the sintering path. In all, the toughness remains low and the brittleness, represented by the index $BI = \frac{H}{K_{1c}}$, high (**Figure 3C**). As expected in ceramics and highly mineralized composites, higher properties could be obtained at finer grain sizes.

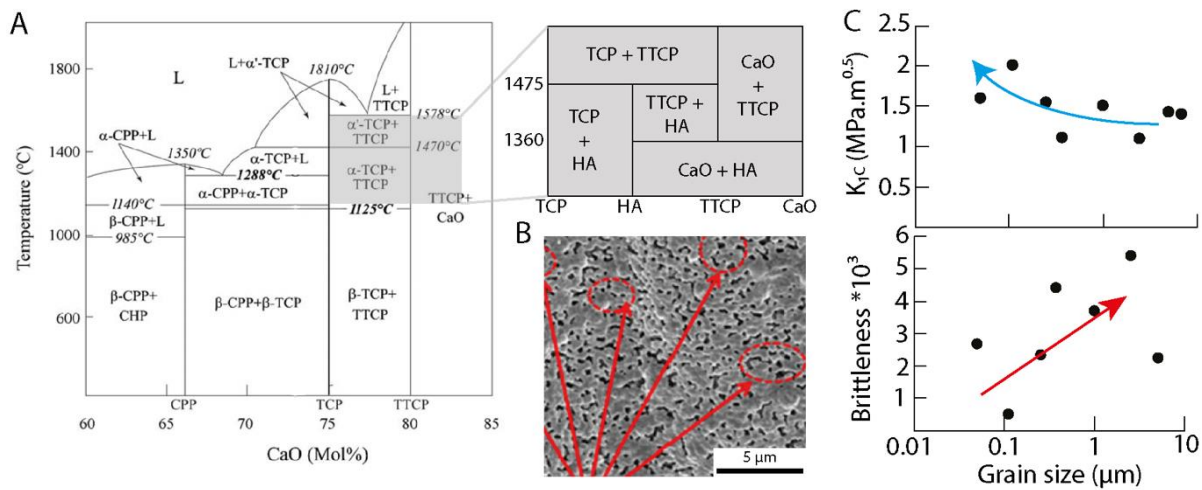

**Figure 3: Polycrystalline CaP materials. (A)** Partial phase diagram of CPCs. The grey insert includes HA phase and corresponds to the phase diagram in presence of water vapor at 500 mmHg pressure. From refs [19,41]. Reproduced with permissions from the Royal Society and from Elsevier. **(B)** Electron micrograph of a sintered ceramic using bone nanoparticles from biowaste. The red lines indicate the presence of nanopores. Reprinted from ref [38] with permission from Elsevier. **(C,D)** Toughness and brittleness index as functions of the grain size in sintered CPCs. Data points extracted from [20,22,26,28,29,38–40].

## 2.2| Microscale: Adding porosity

Although dense ceramics are best for high static mechanical performance, microporosity is also important the osseointegration of the implant. Indeed, the presence of micropores in bones, and their associated roughness, increase the surface area available for cell attachment and bone formation. This stabilizes the interface between the implant and the remaining bone, reducing micromotion and the associated inflammation. Furthermore, the osteons in the cortical bone, large channels in the range of 50 to 90 µm, provide paths for nutrient and waste transport and host blood vessels and nerves [42]. Finally, porosity associated to the cancellous part of the bone decreases the overall weight. A porosity of 40 to 60% has been found to be ideal for the bioactivity of a porous implant [43]. Although porous scaffolds are generally associated with TE approaches and weak materials, porosity can be implemented into CPCs using various techniques to tune the mechanical properties (**Figure 4**).

Methods to create porous microstructures include foam replication where a sponge is coated with a ceramic slip followed by its burning out [44], the use of sacrificial porogen particles [45], gel casting [46] and freeze-casting [47,48]. Freeze-casting is a method of choice for CPCs due to the formation of large and elongated interconnected channels and the large tunability in pore sizes. During freeze-casting, a ceramic slip is gradually frozen by contact with a cold source. After freezing, the ice is removed by sublimation to reveal the channels before being sintered (**Figure 4A**). The freezing conditions control the pore size while the morphology remains lamellar due to the formation of ice dendrites. For example, freezing a slurry containing 40 vol% of β-TCP microparticles at a constant freezing rate (CFR) of 1.86 °C/min results in a pore size of 4.83 μm as compared to 2.84 μm under constant freezing temperature (CFT) of 5 °C [48]. As a consequence, the compressive strength of the CPC made under CFR exhibited a lower compression strength of 1.74 MPa as compared to 2.25 MPa under CFT conditions [48]. Furthermore, the pore sizes also depend on the dimensions of the initial powders. Nanopowders can achieve large pore sizes of ~100 μm as they can pack densely between the ice dendrites. Conversely, freezing slurries containing large micropowders leads to smaller pore sizes below 10 μm.

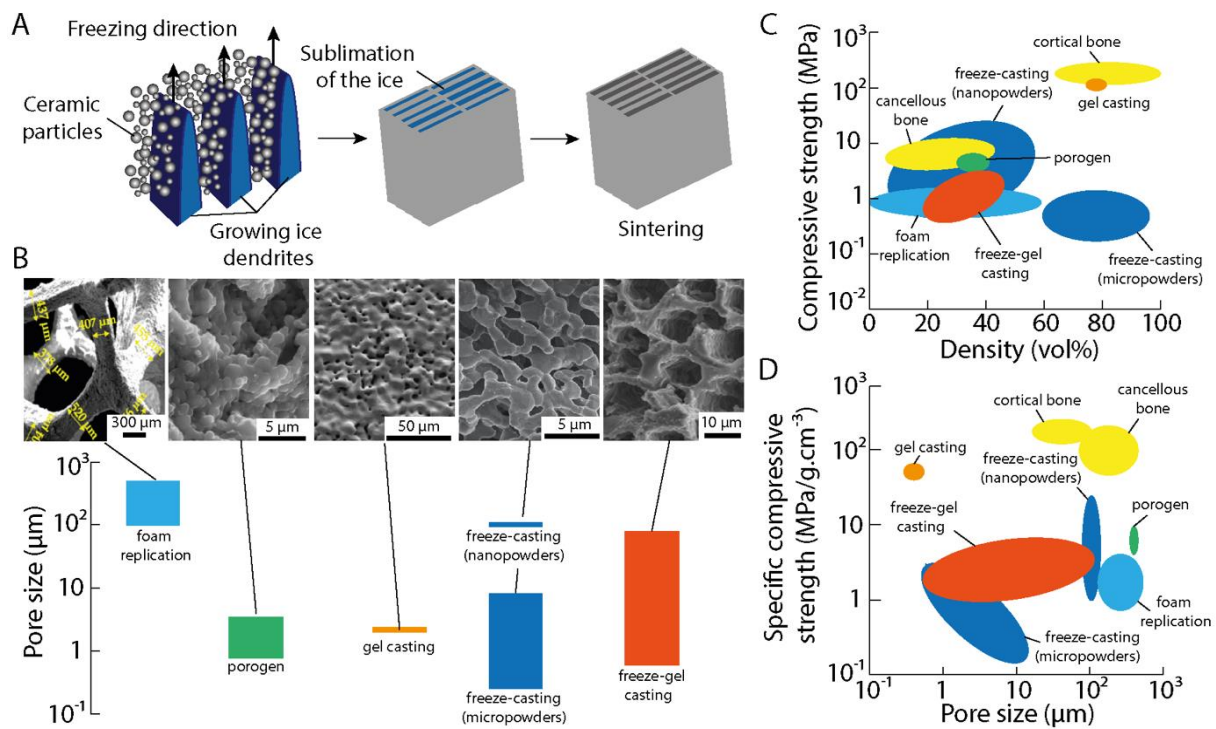

**Figure 4: Microscopic porosity in CPCs: (A)** Schematics of the freeze-casting process. **(B)** Plot and electron micrographs comparing the pore size range achieved in CPCs made using foam replication (light blue), porogen (green), gel casting (light orange), freeze-casting (dark blue), and freeze-gel casting (dark orange). Data from ref [47–53]. Micrographs reprinted from [48,50–52,54] with permissions from Elsevier. **(C,D)** Compressive strength as a function of density and specific compressive strength as a function of pore size for porous CPCs and bone.

To increase the mechanical strength after sintering, freeze-casting has been combined with gel casting. In gel casting, the ceramic particles are suspended in an organic monomer solution that is then polymerized [46,51]. In freeze-gel casting, the gel formation inhibits the growth of ice dendrites, creating irregular pore channels instead of lamellae [52,55]. After sintering, the ceramic thus contains a bimodal pore distribution with pores of ~1 µm left by the polymer and pores of ~ 80 µm left by the ice [48]. This wide range of pore sizes contrasts with gel casting alone that only yields submicrometric pores [48], or foam replication that only produces large pores >100 µm [53] (**Figure 4B**). However, gel casting remains the method providing the highest compressive strength in CPCs, up to 130 MPa [51] as compared to only 4 MPa with freeze-gel casting (**Figure 4C**). This high compressive strength is presumably due to the high relative density of 80% and small pore sizes in gel casted CPCs [46]. Gel casting thus leads to compressive strengths close to that of cortical bone, whereas the other methods, in particular freeze-casting of nanopowders and the use of porogens lead to mechanical properties close to that of cancellous bone. However, comparing the specific compressive strengths and pore sizes of porous CPCs and bone highlights that gel cast ceramics are not the best candidates (**Figure 4D**). Instead, freeze-casting using slurries of nanopowders can achieve high specific compressive strengths for pore sizes similar to that of bones. However, those porous ceramics still remain brittle.

**2.3| Macroscale: Designing the shape and geometry**

Controlling the macroscopic shape and geometry of bone implants are also required for efficient load transfer and morphogenic reasons. Indeed, a bone implant with an odd shape could result in bone growth or decay following Wolff's law. Also, the presence of gaps could lead to difficulties to integrate and bind to the remaining bone. Shaping can be realized *via* three-dimensional (3D) printing. Using computer-aided designs to build structures in a layer-by-layer fashion, images aquired by computed tomography or magnetic resonance scans can be used to capture the internal structure and overall shape specific to each patient [56,57]. 3D printing techniques are thus interesting for point-of-care use and patient-specific designs. Four 3D printing methods have been explored to fabricate CPCs. These methods are divided between powder-based and liquid-based methods (**Figure 5**). Each approach follows three main steps: printing of a green body, burning out organic binders and additives, and consolidation by sintering. In this section, we highlight the advantages and drawbacks of the main 3D printing methods applied to CPCs, and compare the mechanical properties achieved.

Powder-based 3D printing methods are powder-3D printing (P-3DP) or binder jetting, and selective laser sintering (SLS) or powder bed fusion (**Figure 5A**). In both methods, each layer is initially made of a bed of ceramic powders. During P-3DP, an organic binder is locally deposited onto the powder bed at the areas desired to be consolidated to form the final object. This step is then repeated for each layer. Before sintering, the excess powder is removed. P-3DP thus allows flexible print designs and complex 3D structures. For example, overhangs of several millimeters can easily be created since the material is always maintained by the powders from the underlying layers [58] (**Figure 5A**). However, the printed materials generally feature high residual porosity due to loose packing and coarse sizes of the powder, as well as the high fraction of binder. Indeed, the organic binder should be sufficient to consolidate and handle the green part before sintering. Also, coarse powders of 30 to 100 µm diameter are generally used to prevent agglomeration [44,59]. Mixtures of coarser and finer powders can be packed tighter in the powder bed, thereby increasing the strength after sintering. For example, by mixing 15 to 25 wt% of finer HA powder of 30 µm diameter to HA powderwith diameters higher than 125 µm, the compressive strength after sintering increased from 10 to 15 MPa [60]. However, the microvoids that remain still greatly weaken the CPCs [59]. SLS is another powder-based method that locally fuses powders with a laser beam (**Figure 5B**). Similarly to P-3DP, SLS allows flexible print designs. However, the thermal shocks that occur during laser binding often lead to internal stresses [56,59]. For example, SLS-printed ß-TCP scaffolds exhibited a toughness lower than 1.5 MPa.m$^{0.5}$ [61]. Likewise, the large powder sizes required for powder-bed flowability limits the resolution.

The other 3D printing methods used for CPCs are liquid-based. Using particles suspended in solvents or polymer solutions permits the use of fine colloidal powders at high concentration and with minimal agglomeration. These features greatly improve the resolution and the surface finish of the printed parts as compared to the powder-based technologies. Direct ink writing (DIW), or robocasting, directly extrudes a ceramic paste or ink through a nozzle to create the green part [56] (**Figure 5C**). This method uses minimal amount of material but has limited design complexity. Logpile structures are the most frequently studied scaffolds [62–64]. Indeed, overhanging struts tend to sag and require a sacrificial supporting material. Furthermore, the ink needs to be optimized to exhibit shear-thinning and thixotropy for extrudability and shape-retention. Tunability can nevertheless be added by exploring nozzle shapes and ink compositions. For instance, coaxial needles can extrude hollow struts [65] (**Figure 5C**). Water-based ceramic ink with 45 vol% HA was printed through the outer channel of a coaxial needle at 40 mm/s while sacrificial paraffin oil was co-extruded through the inner channel at the same rate. This way, the HA ceramic created a strong and stiff shell after burnout and sintering [65]. Another common liquid-based 3D printing method is vat photopolymerization, which includes as stereolithoghraphy (SLA) annd digital light processing (DLP) (**Figure 5D**). In this process, the powders are suspended in a photocurable resin and cured by UV light [44]. DLP is a faster variation of SLA where photocuring is done layer-by-layer instead of point-by-point [43]. Complex print designs are possible but light scattering limits the resolution due to the difference in optical density between the ceramic particles and the resin. Ridges caused by line broadening effects of the UV light are often seen [66] (**Figure 5D**).

Comparing the four 3D printing methods, a trade-off has to be made between resolution and printing speed (**Figure 5E**). Vat photopolymerization provides the highest resolution but the printing speed is the lowest. On the contrary, P-3DP has the highest printing speed but a low resolution. As a compromise, DIW can achieve high printing speed with micrometric resolution. However, the design freedom is limited by the rheological requirements of the inks and sagging of overhangs. Nevertherless, 3D printing enables macroscopic shapes in CPCs, at densities varying between 20 and 80%, and with controlled pore sizes from 10 μm to a few mm. Despite design freedom on the macroscale, the mechanical properties in 3D printed CPCs are still poor, with a low specific compressive strength and lack of toughness (**Figure 7F,G**).

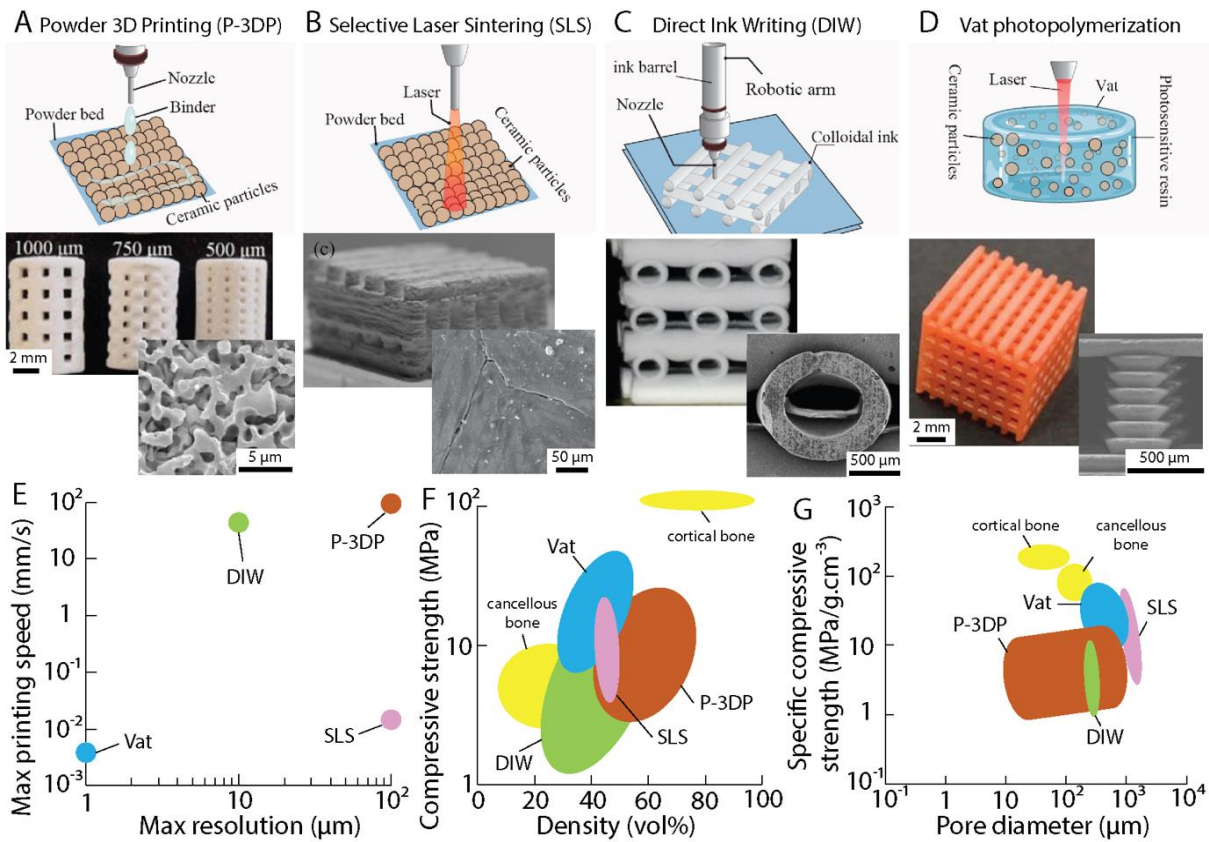

**Figure 5: Macroscopic design using additive manufacturing. (A)** Powder 3D printing and printed sintered TCP. The micrographs highlight residual porosity. Reprinted from ref [58] with permission from John Wiley and Sons. **(B)** Selective laser sintering and printed ß-TCP. The micrograph is a close-up view showing showing a crack that developed during the processing. Reprinted from ref [61] with permission from Elsevier. **(C)** Direct ink writing, picture of a logpile sample with hollow struts and micrograph of the struts' cross-section. Reprinted from ref [65] with permission from Elsevier. **(D)** Vat photopolymerization, picture of a sample made by DLP and close-up view of the line broadening effect caused by light scattering. Reprinted from ref [66] with permission from Elsevier. All schematics in A-D are reproduced from ref [56] with permission from Elsevier. **(E-G)** Maps representing the printing speed as a function of the resolution, compressive strength as a function of the density, and specific strength as a function of the pore diameter, for CPCs obtained using the four 3D printing methods and for natural bone.

Existing methods to control the crystal phases, grain sizes, micropores and macroscopic geometries in CPCs have provided a large body of research. However, the mechanical properties and in particular the toughness has not been achieved by tuning the microstructure at selected lengthscales. Inspired by natural materials, researchers have put efforts in creating complex structures combining controls at multiple lengthscales simultaneously [18]. These are reinforced microstructures and multi-level hierarchical structures. With such materials, the level of complexity and microstructural design is increased, to better reproduce the intricate organisation and hopefully the outstanding properties of natural materials. To clarify the terms, a multi-level structure relates to the juxtaposition of a material with a controlled nanostructure with another materials with a controlled microstructure or macrostructure, for example. In a three-level hierarchical structure, the macrostructured material is composed of a microstructured material, which is in turn composed of a nanostructured material. In the following, methods to create hierarchical structures in CPCs and their composites are described and their properties discussed.

## 3| Hierarchical bioinspired approaches

### 3.1| Reinforced microstructures in CPCs and highly mineralized composites

As a first approach, reinforcing dense CPCs with nanofibers, nanoplatelets, or nanoparticles has been explored (**Figure 6**). Generally, the use of nanoelements in a microstructure increases the toughness *via* extrinsic toughening mechanisms such as crack bridging, crack deflection and twisting, and reinforcement pull-out, similarly to what is observed in cortical bones (**Figure 2B-D**).

HA ceramics have been reinforced with nanoparticles that have some intrinsic toughness, like $ZrO_2$ [67–69]. The toughness of the HA-$ZrO_2$ ceramic composites increased from 0.8 to 2.3 $MPa.m^{0.5}$ as the content in $ZrO_2$ increased [70]. Higher toughness could be reached with the addition of 20 wt% of $SiO_2$ nanoparticles to the $ZrO_2$-HA composition [69]. Indeed, $SiO_2$ produced 99% densification after the pressureless sintering at 1200 °C, while maintaining nanograins [69]. In absence of $SiO_2$, the grains after sintering were of 1 to 5 μm. Although $ZrO_2$ is also a biocompatible ceramic, the HA phase decomposed into $\alpha$ and ß-TCP, as well as phases like $CaZrO_3$ or $Ca_2ZrSi_4O_{12}$. MgO has also been used to promote the pressureless densification of HA but yielded microsized grains [71]. To maintain the HA phase and nanosizes throughout the process, 40 wt% $CaTiO_3$ particles have been added to HA and sintered at 1200 °C under pressure [72]. However, the 98% dense ceramic obtained had a toughness of 1.7 $MPa.m^{0.5}$ only. The mechanical properties of current reinforced CPCs are higher than pure CPCs like HA, exhibit biocompatibility, but are still lower than cortical bones (**Figure 6A**).

1D and 2D nanomaterials with high aspect ratios are other ways to reinforce HA ceramics while maintaining the crystal phase and biocompatibility. Examples of such nanomaterials are BN nanotubes (BNT), carbon nanotubes (CNT) and graphene. For example, adding 4 wt% of BNTs of aspect ratio of ~50 dispersed with HA nanorods and sintered under pressure at 1100 °C lead to a 97% dense ceramic with nanometric grains and a toughness of 1.6 $MPa.m^{0.5}$ [73]. The ceramic also had a high wear resistance and was viable for osteoblasts [73]. Addition of CNTs and graphene also led to increasing toughness up to 2.4-2.5 $MPa.m^{0.5}$ [27,74] (**Figure 6B**). One interesting aspect of using anisotropic nanoreinforcements is the possibility to create anisotropic properties, for example using pressure-driven alignment [74,75]. Alignment increases pull-out mechanisms and imitate the structural anisotropy of bones.

Anisotropy in bulk CPCs can also be tuned in any direction by using a high strength external magnetic field [76–81], with ß-TCP and HA as initial powders. In the process, the slurries were casted onto porous moulds under a magnetic field of 4 to 10 T. Rotation of the sample was applied to increase the magnetically-directed alignment of the particles. After the casting and drying, pressureless sintering lead to abnormal grain growth leading to anisotropic and oriented grains [78] (**Figure 6C**). However, although the ceramics obtained were 99% dense and exhibited a well-defined anisotropic texture, there is no evidence of increase in toughness or flexural strength. The large grains from 4 μm to 50 μm presumably lead to rather brittle materials.

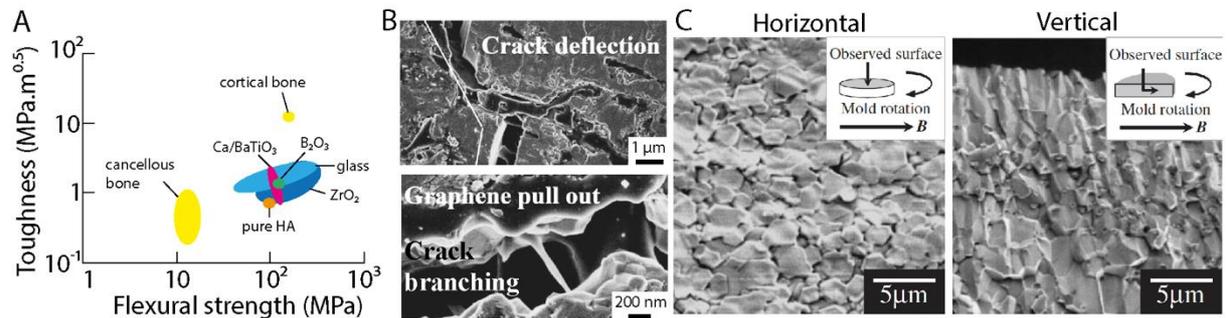

**Figure 6: Reinforced CaP materials: (A)** Toughness as a function of the flexural strength for dense ceramics reinforced with $ZrO_2$, $CaTiO_3$ or $BaTiO_3$, $B_2O_3$ particles, and glass phases. Data extracted from refs [69,72]. **(B)** Electron micrographs of a CPC reinforced with graphene, showing crack deflection (top) and graphene pull-out and crack bridging (bottom). Reprinted from ref [27] with permission from Elsevier. **(C)** Micrographs of textured HA ceramics prepared by magnetic slip-casting showing horizontal and vertically oriented grains. Reprinted from ref [79] with permission from The Japan Institute of Metals and Materials.

### 3.2| Multi-level hierarchical structures inspired by bones

Reinforcing matrices with nanoparticles could moderately increase the toughness of CPCs without compromising on the flexural strength. More complex hierarchical structures inspired by the organization of bones have been created to further improve the fatigue resistance of CPCs. These hierarchial structures intend to mimic the macroscopic organization of long bones with the porous zone enclosed and protected by a dense shell, the presence of organic molecules for binding and toughening the ceramic structure, and biomimetic mineralization to create highly mineralized microstructures (**Figure 7**).

Macroscopic organisation with a dense and a porous part can be achieved using a laser beam to melt the outer surface of porous freeze-casted HA scaffolds [82]. Another method uses lost-wax molding to create a porous β-TCP with a dense outer shell. The compressive strength of the overall structure increased from 9 to 36 MPa with the addition of the dense shell [83] (**Figure 7A**). To create graded variations in porosity, gel casting and tape casting have been combined to deposit layers of BCP slurries containing polymerizable monomers that simultaneously served as binder and porogens [84]. No increase in mechanical properties was reported but the method demonstrates the feasibility to tune the local porosity in a macroscopic ceramic.

In addition, bones are composite materials containing a large fraction of CaP nanoparticles interacting with biomolecules like collagen. Porous CPCs can be infiltrated by polymers. For example, gelatin can be infiltrated in micropores and leave a coating on 3D printed struts that maintain the cohesion of the material, even after fracture [45]. The compressive strength was also found to double [45]. In another example, a freeze-casted scaffold containing 84 vol% of HA platelets was infiltrated with poly(methyl methacylate) (PMMA), increasing the flexural strength up to 119.7 MPa [85]. Further grafting HA with 3-(trimethoxysilyl)propyl methacrylate prior to the infiltration of PMMA and polymerization ensured a strong HA-PMMA interface improving stress transfer and tearing the polymer under strain [85]. In this configuration, the polymer phase can contribute to the energy absorption and the toughness of the composite. In another study, a CaP scaffold obtained by foaming a slurry of HA and β-TCP was infiltrated with collagen and HA nanoparticles were synthesized directly inside the collagen matrix [86]. The overall hierarchical structure exhibited a 3 fold increase in compressive strength as compared to the CaP scaffold alone, from 2 MPa to 6 MPa [86]. Thanks to this multi-level hierarchical organisation, the inorganic and organic components interacted synergistically. As the collagen held the HA nanocrystals together, those nanocrystals reinforced the collagen matrix and in turn the overall CaP scaffold.

Finally, biomimetic mineralization can be used to create hierarchical CPCs. This was explored using a bottom-up crystal construction where HA nanorods were grown onto pre-synthesized DCP microplatelets [87]. After infiltration with gelatin, the composites exhibited an elastic modulus of 25.9 GPa and a hardness of 0.90 GPa, which are comparable to cortical bone [87]. A faster and more scalable approach employed a hydrothermal treatment to convert α-TCP powders into calcium-deficient HA (CDHA) nano-needles [88]. Those CDHA needles were densely packed into bundles stacked in random orientations, deflecting cracks around each bundles **(Figure 7B)**. The hierarchical structure exhibited a 5-fold increase in the work of fracture, while maintaining 35% porosity [88].

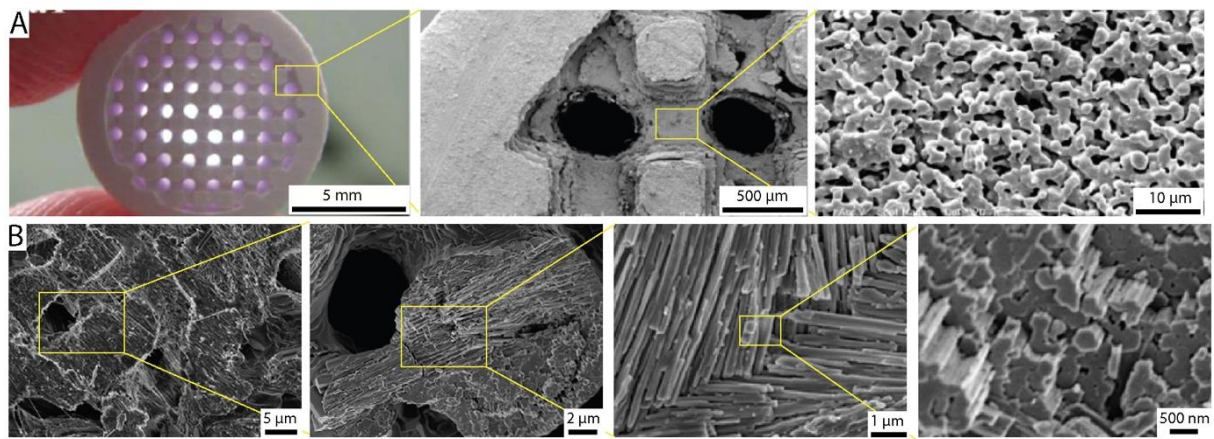

**Figure 7: Hierarchical CPCs inspired by bones. (A)** Electron micrographs of CPCs with hierarchical porosity: a dense shell, microchannels and micropores. Reprinted from ref [83] with permission from John Wiley and Sons. **(B)** Micrographs showing a hierarchical organisation obtained *via* biomimetic mineralization and showing pores and CaP microfibers in aligned bundles oriented in multiple directions. Reprinted from ref [88] with permission from Elsevier.

The multi-level hierarchical structures inspired from bone explored changes in density, the incorporation of ductile and absorbing polymers, and biomineralization of aligned crystalline bundles. These strategies seem to improve fracture properties of CPCs, although toughness values are not often reported. Furthermore, despite these methods draw inspiration for the hierarchical structure of bones, they stay rather simplistic as compared to the natural material due to technical limitations. Several other strategies have been explored to increase this level of complexity, by taking inspiration from other biological materials.

**3.3| Multi-level hierarchical structures inspired by other natural materials**

Multi-level hierarchical structures made of CaP have been created using inspiration from other tough biomaterials such as wood, seashells, and arthropods' exocuticles. Biotemplating methods, brick-and-mortar arrangements, and twisted plywood structures have been applied to CaP materials to increase their toughness (**Figure 8**).

Biotemplating is method where a hydrothermal treatment converts a biological material into HA while retaining the natural porous and complex nanofeatures of the original template (**Figure 8A**). For example, biotemplating of corals reproduced the nanopores of 5 to 50 nm and the interconnected network of macropores of 150 to 500 µm [89] (**Figure 8A, top**). In another example, decellularised cuttlefish bone was converted into HA, then coated with a mixture of polycaprolactone and polylactic acid [91] (**Figure 8A, middle**). The resulting Young's modulus reached 12.1 MPa, 18 times higher than that of the original cuttlefish bone while the porous hierarchical microstructure was preserved [91]. Similarly, biotemplating of Rattan wood into HA reproduced the interconnected porous network formed by the veins of the plant [92] (**Figure 8A, bottom**). These ceramics remain brittle but the method could be augmented with the strategies described in previous sections 3.2 to increase their toughness.

Besides porous materials, the brick-and-mortar arrangement of the nacreous layer of seashells has been widely used as inspiration for strong and tough ceramics and composites (**Figure 8B**). Layer-by-layer methods or vacuum- assisted deposition could reproduce this type of structure in CaP materials in a scalable fashion [93,94]. For example, using repeated lamination, bulk DCPD-sodium alginate composites with dimensions of 10×10×0.5 cm$^3$ were made [93]. A solution containing chitosan and Ca$^{2+}$ was added between each DCPD layer to prevent delamination. The final lay-up was then hot-pressed to consolidate the structure and increase the mineral concentration of minerals to 50 wt%. The resulting composite had a flexural strength reaching 175 MPa and exhibited resistance to crack growth with a $K_{JC}$ increasing from 3 to 8 MPa.m$^{0.5}$ as a crack propagated. Microscopic investigations of damaged samples revealed a tortuous crack path with microcraking and multiple deflections (**Figure 8C**). In another nacre-like structure, HA platelets were oriented in a matrix composed of amyloid fibrils at a mineral content of 40 wt%. Although the fracture behaviour of the material was not reported, the modulus reached 1.3 GPa, similar to the highest level of cancellous bones, while the overall composite had very good osteogenicity [94].

Finally, another natural highly mineralized composite that exhibits outstanding strength and toughness is the twisted plywood or Bouligand structure found in the exocuticles of stomatopods and arthropods [95]. This type of arrangement could be reproduced in CaP materials using a brush deposition method [96] (**Figure 8D**). Typically, slurries containing 40 wt% of HA microfibers suspended in a sodium alginate solution were deposited using a brush. The motion of the brush led to the shear-induced orientation of the HA microfibers. By brushing multiple layers with various shear directions, twisted plywood structures were constructed with controlled angles between the HA orientations in each layer. The flexural strength of the resulting composites decreased as the angle between consecutive layers increased, from nearly 275 MPa at 0° angle to 175 MPa at 90°. However, all samples exhibited toughness and a rising R-curve. In particular, 10° angle between consecutive layers exhibited the highest toughness with a stress intensity factor rising from 3 to 8 MPa.m$^{0.5}$ as a crack propagated. X-ray computed tomography and fractography showed that crack deflection and crack twisting were the prevalent toughening mechanisms.

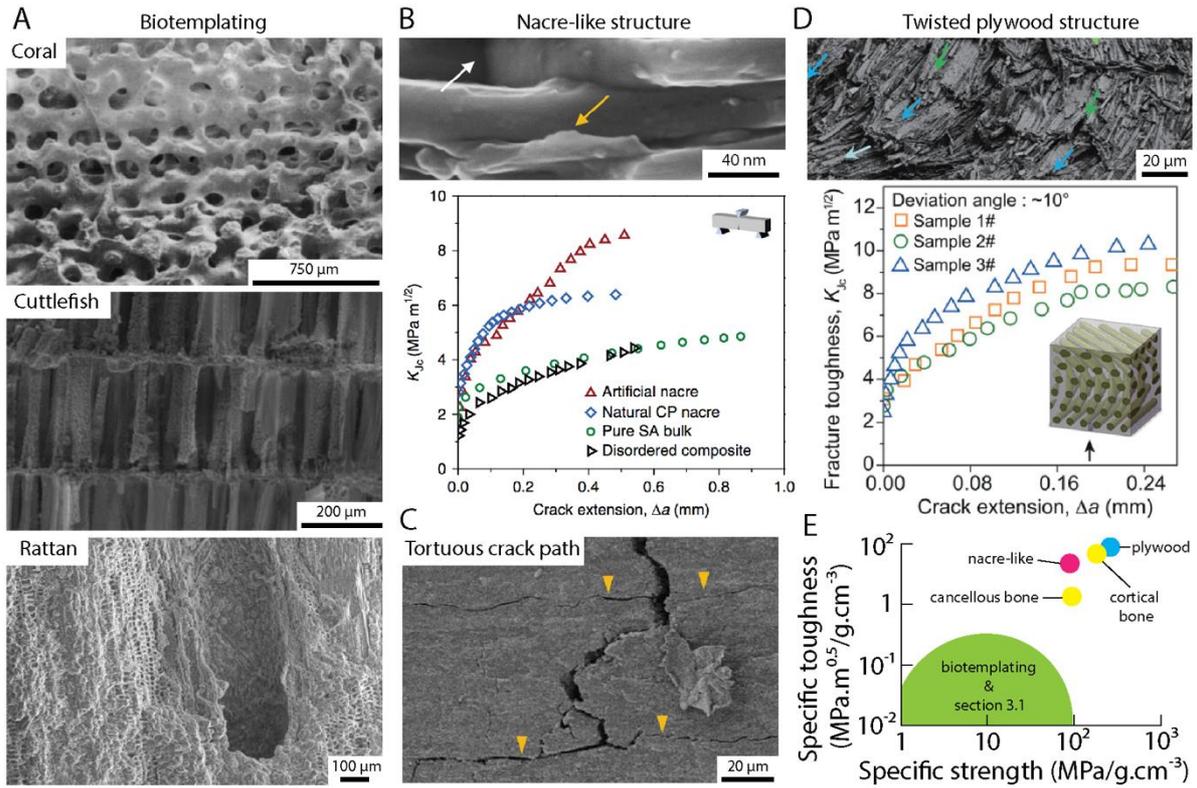

**Figure 8: Hierarchical CPCs inspired by other natural materials. (A)** Micrographs of CPCs obtained by biotemplating corals, cuttlefish and rattan wood. Reprinted from [89] with permission from Elsevier. Reprinted from [91] with permission from John Wiley and Sons. Reprinted from [92] with permission from Elsevier. **(B)** Micrograph of a HA-sodium alginate (SA) composite inspired by nacre and the corresponding crack resistance curve (artificial nacre, triangle). Natural CP nacre refers to the *Cristaria plicata* seashell, disordered composite corresponds to a HA-SA material of same composition without the brick-and-mortar structure, and pure SA bulk refers to the pure SA matrix. Reprinted from ref [93] with permission from Springer Nature. **(C)** Micrograph of the crack path in the artificial nacre. Reproduced with permission from ref [93] with permission from Springer Nature. **(D)** Micrograph of the structure of a twisted plywood composite made of HA microfibers and SA, and the corresponding resistance curve, for an angle of 10° between each layer. Reprinted from ref [96] with permission from China Science Publishing & Media Ltd. **(E)** Property map comparing the specific toughness and strength of hierarchical CaP materials inspired by bones or other natural structures. The green area reports brittle CaP materials obtained via biotemplating and the approaches described in section 3.1, pink is the artificial nacre, blue the artificial plywood and in yellow the natural bone.

Comparing the specific strength and the toughness of CaP materials inspired from complex hierarchical biomaterials, it appears that twisted plywood structures are performing the best (**Figure 8E**). However, it is difficult to compare directly methods that employed raw materials of different chemistries but most methods do not achieve the specific toughness of cortical bones. Nacre-like structures in CaPs have a high specific toughness but the specific strength achieved is closer to that of cancellous bones. In addition, the advantage of the brush method used for the twisted plywood as compared to biotemplating or vacuum-assisted deposition lies in its simplicity, high structural control, and adapatability to surfaces of various shapes. However, these structures lack interconnected pores or long channels for vascularization. There could thus be a benefit in combining the different approaches to reproduce the local organization of bones.

## 4| Limitations, challenges, and suggestions

Despite the large amount of research produced to fabricate materials for bone implants, further research is still needed to match the properties and functionalities of bones. There are indeed multiple challenges to face to overcome the current limitations in mechanical properties and biological response of today's implants, namely poor specific strength and toughness, Young's modulus mismatch, and lack of integration with the surrounding tissues and the vascular and nervous systems. However, thanks to the recent advances in the fields of additive manufacturing and bioinspired materials, future prospects are promising. In the following are summarized the most important challenges and selected works on the fabrication of other ceramics and composites that could disrupt and boost the development of more suitable bone implants.

### 4.1| Challenges

There are three main challenges hindering the fabrication of load-bearing bone implants with properties similar to those of natural bones: a lack of communication and collaboration between research communities, incomplete understanding of the natural bone composition, microstructure, and evolution, and the lack of control in the sintering of CPCs.

First, the development of bone implants has been enterprised by two communities in parallel, the biomedical engineers, in particular the tissue engineering community, and the materials scientists. This disjunction has led to a missing overarching view of the technical issues as well as incomplete data sets. On one side, biologists focused on the biological properties, thus primarily designing porous structures of controlled chemistry. On the other side, materials scientists aimed at achieveing the mechanical properties without necessarily considering the biological response and the suitability of materials for clinical applications. Luckily, the past decade has seen many multidisciplinary teams blossom, where engineers are bridging the different disciplines. Engineers are aiming at fulfilling both biological and mechanical requirements with a clear and practical goal, that of using the final structure for bone repair.

The second challenge to face is the still incomplete understanding of the structure and remodelling of our bones. Despite impressive recent results [14], new insights on bone structural arrangement [97] and crystalline phases [98] are regularly discovered as more advanced characterization methods develop. Transposing the current understanding to engineering is scaling up the difficulty. For example, biological CPCs are formed in the body under mild conditions through protein-guided biomineralization. Although the biomimetic mineralization of CaP particles of different shapes has been explored, they tackle mostly nanospheres, one-dimensional rods, whiskers and wires, two-dimensional platelets or sheets, hollow microspheres for drug loading applications, or enamel-like coatings [99]. The natural processes that lead to biomineralization and their assembly into complex structures are still hypothetical [100–102]. In practice, this challenge has been well exemplified with the case of nacre, where a demineralized seashell scaffold was incubated with amorphous nanoparticles and polyaspartic acid. The resulting material had a microstructure similar to the natural nacre, but the remineralized nacre consisted of calcium carbonate in the calcite phase instead of the natural aragonite phase, leading to a lower hardness [103]. This example shows how biomineralization processes need to be better understood for their direct transposition into artificial hierarchical materials.

Finally, the third main challenge is to better study the theoretical and practical effects of sintering of CPCs, during which shrinkage and phase transformations occur. The diffusion process of sintering, burnout of organics and crystal changes are likely the cause of residual stresses after sintering, weakening the material. In addition, sintering leads to grain growth that affects the hardness, strength, toughness, as well as the biological response. During heat treatment, the interfacial properties between the grains are also tuned. These interfacial properties are then controlling the fracture path and energy in ceramics and composites [104,105]. Lack of knowledge about the interfacial properties thus limits the development of materials where cracks and deformations can be directed and controlled.

Addressing these challenges may give insights on the ideal designs to use for bone repair. Notably, these are general scientific questions that also need to be addressed for other materials compositions. The advancing of characterisation techniques such as *in situ* confocal Raman microscopy, liquid Transmission Electron Microscopy and virtual mechanical testing could provide answers to these questions and suggest new strategies [106–108].

## 4.2| Future prospects

Despite these challenges, many recent papers are describing promising avenues towards more biocompatible processes, with more design flexibility and with scalable, customizable and adaptable capabilities. Biocompatible methods have been developed to increase the bioactivity of 3D printed parts through the incorporation of biomolecules such as growth factors. For example, low-temperature P-3DP uses phosphoric acid as a binder to strengthen parts by dissolution-precipitation without producing excessive acidity [109–111]. In vat photopolymerization methods, nontoxic monomers such as polyethylene glycol diacrylate can be used as resin [112]. In the following are presented a few selected works that are promising for the scalable fabrication of customized and performant load-bearing bone implants (**Figure 9**). These processes have been developed for other chemical compositions and could be adapted to CaP materials.

First, building materials *via* 3D printing is convenient to control both the macrostructure and the microstructure of materials. One way to control the local microstructure is to combine 3D printing with external fields like acoustic, electric or magnetic [113]. This allows to build anisotropic and oriented microstructures locally into an overall complex design. One particularly interesting example of local controlled orientation in a dense composite was made using DIW of an ink containing $Al_2O_3$ microplatelets [114]. The shear stresses that developed in the nozzle during extrusion oriented the microplatelets in concentric orientations [114] (**Figure 9A, top**). After printing a macroscopic twisted plywood architecture, the sintered part was infiltrated with an epoxy and tested for crack propagation using single edge notch beams. A rising R-curve was recorded with a stress intensity factor increasing from 3 to 9 $MPa.m^{0.5}$. This toughness is related to the tortuous crack path that developed in the material (**Figure 9A, bottom**). This method is promising to build local orientation patterns in pre-determined shapes and porosity.

Furthermore, layer-by-layer deposition can be carried out using traditional ceramic processing methods like slip casting, and augmented with an external magnetic field to control the local orientation. Alumina composites and ceramics with nacre-like structures showing strength and toughness have been realised using this method [115]. Furthermore, the combination of externally-driven colloidal assembly during the casting process allowed the fabrication of programmed local microparticles orientations in particulate composites and ceramics [115,116]. Using the method, a biomimetic ceramic tooth with a bilayer architecture was made, that combined toughness, strength, and a gradient in hardness [115] (**Figure 9B**). The tooth-like ceramic, made of $Al_2O_3$ microplatelets, had vertical and horizontal grain orientations (**Figure 9B, top**). The horizontal alignment recalled the brick-and-mortar structure of nacre that exhibits high crack resistance. In addition, similarly to the dentin and enamel layers, the outer part of the sample had a higher density with the addition of silica nanoparticles. After sintering, a crack propagating inwards is deflected at the interface between the two layers (**Figure 9B, bottom**). One advantage of the approach lies in its adaptability, simplicity, and tunability. Particles of other dimensions and chemistry can be manipulated after being functionalized with magnetic nanoparticles.

Finally, an interesting approach consists in using top-down methods like laser engraving and cutting. Indeed, a brittle material can be toughened by carefully creating weak interfaces of controlled shapes and geometry. This method was applied to transparent silica glasses [117,118]. In the process, alternating layers of laser-cut sheets of glass and elastomers were stacked. Building a twisted arrangement with an angle of 25.7° between each layer of laser-engraved glass, the damage zone size upon impact was found to be ~50% larger that stacked uncut glass, indicating larger stress dissipation across the material. Another similar approach exploits the friction between unit blocks of defined geometry to dissipate energy. The structures made using these blocks are known as topologically interlocking materials (TIMs) [119,120] (**Figure 9C**). The key advantage of TIMs is that no elastomeric layer is required to increase the flexural strength and impact resistance of panel made from ceramic blocks. Indeed, in TIMs, repeating units of blocks are arranged in a stable periodic configuration that resists translation and rotation. The blocks may be prisms, polyhedrons, or geometries with non-planar faces (**Figure 9C, top**). The assembled blocks are held together in a rigid frame with some pre-stress from the overall frame or another strategy [119]. Upon loading perpendicularly to the panel, the individual blocks can be displaced but the panel does not fracture (**Figure 9C, bottom**). This toughening effect can be tuned by controlling the geometry of the blocks and thus their friction. This strategy could be applied at different lengthscales and for large variety of blocks, which can be fragile like glass or be microstructured [121,122].

3D printing combined with external fields, magnetically assisted slip casting and interlocking structures are thus other approaches that were successfully applied to toughen ceramics. Their application to CPCs for bone repair applications remain to be explored.

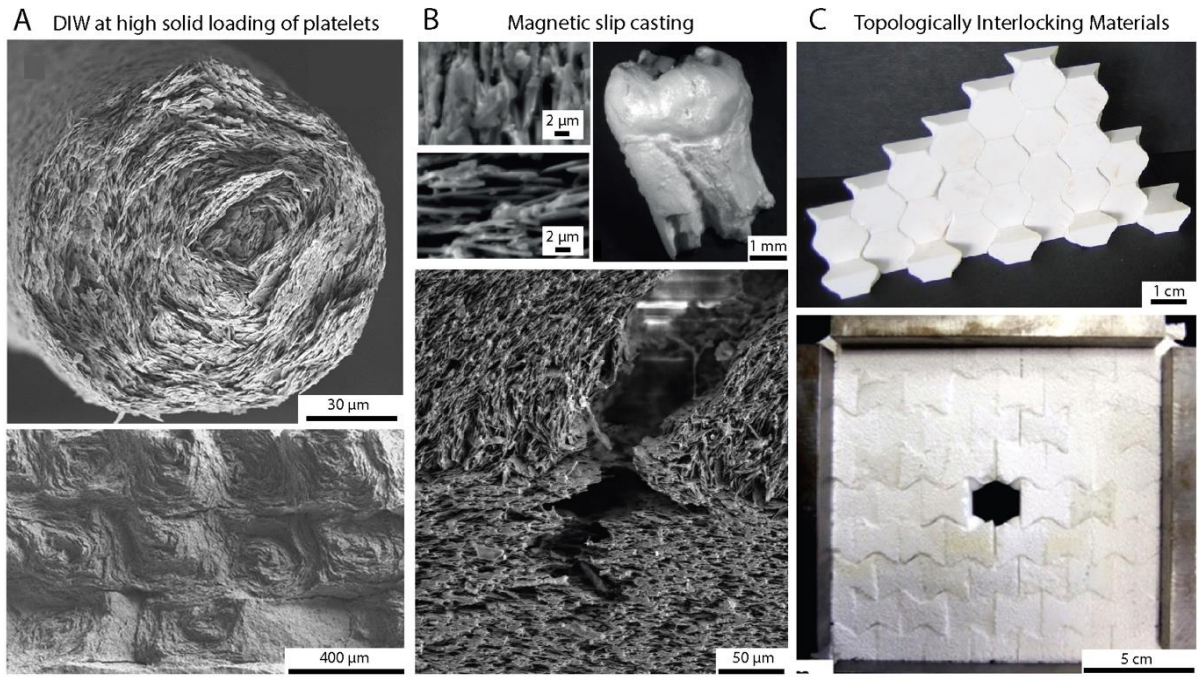

**Figure 9: Recent advances in the development of strong and tough ceramics: (A)** Using DIW for inks with high solid loadings in $Al_2O_3$ microplatelets: top, electron micrograph of a filament extruded from a 25 mm long nozzle, showing the concentric orientation of the microplatelets; and bottom, electron micrograph of a macroscopic twisted plywood structure after sintering and fracture, showing non-brittle fracture. Reprinted from ref [114] with permission from Springer Nature. **(B)** Using magnetically assisted slip casting on slurries containing $Al_2O_3$ microplatelets: top electron micrographs showing the vertical and horizontal platelet alignment and the overall tooth shape of the sintered bioinspired ceramic; and bottom, micrograph showing a crack propagating in the structure. Reproduced with permission from ref [115] with permission from Springer Nature. **(C)** Using topologically interlocking materials with mullite-$Al_2O_3$ blocks: top, assembly of the blocks; and bottom, picture of a topologically interlocking panel that has fractured only locally under a load applied at its centre. Reprinted from ref [123] with permission from Elsevier.

## 5| Conclusion

We have reviewed the fabrication methods and strategies to fabricate CPCs for bone repair (**Figure 10**). CPCs and their composites can be processed using traditional ceramic methods to tune the mineral phases and grain sizes to present stiffness and bioactivity, microporosity can be added *via* a range of strategies comprising foaming and freezing methods to yield light materials, and macroscopic shapes and scalable designs can be realized by 3D printing. The CPCs resulting from these methods nevertheless lack in toughness, which is detrimental for their long term use and performance as load-bearing implants. Inspired by the complex microstructural and hierarchical arrangements of natural biomaterials, multi-level hierarchical designs have been proposed. Reinforcing composites and ceramics with anisotropic nanoparticles allows anisotropic properties and the development of extrinsic toughening mechanisms. Reproducing the organization of bones with local porosity, soft and hard components, and local orientation of minerals provides more complexity to the materials fabricated and increases the energy absorption during fracture. Taking inspiration from the microstructures of other biological materials has led to CaP materials with crack growth resistance. These studies demonstrate the potential of bioinspired approaches to engineering more performant materials that combine multiple functionalities and properties. Also, they highlight the limited technological capabilities to directly copy natural materials. Instead, engineers can build upon the fundamental understanding of the toughening mechanisms in a variety of biomaterials to transpose those mechanisms to artificial engineered materials. Future research in the area of bone implants and bone repair can therefore be expected to emerge from advances in multiple disciplines, namely characterization methods, biology, materials and engineering. Several recent strategies used to toughen brittle materials have been highlighted at the end of this review to suggest next directions of research. In particular, the study of biological materials and methods to transpose their designs to ceramics and composites is anticipated to bring breakthrough in a large range of applications, beyond the biomedical area.

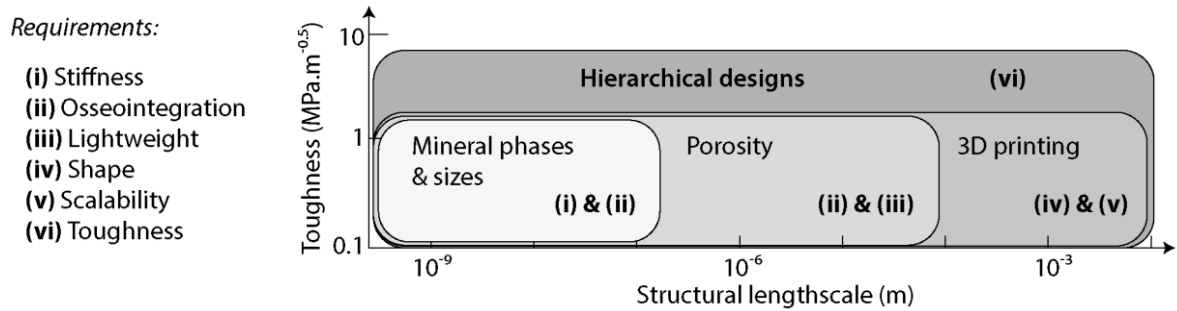

**Figure 10:** List of the six main requirements that materials have to fulfil for their use as successful bone implants and plot showing the toughness as a function of the microstructural dimension to realize them.

**Acknowledgments**
The authors acknowledge financial support from the National Research Foundation, Singapore (Fellowship NRFF12-2020-0002).

**Competing interests**
The authors declare no competing interest.

**Credit author statement**
Peifang Dee: Conceptualization, Data Curation, Writing.
Ha Young You: Data Curation, Writing.
Swee-Hin Teoh: Reviewing and Editing.
Hortense Le Ferrand: Conceptualization, Supervision, Writing, Reviewing and Editing.

**Graphical abstract:**